\documentclass[draftclsnofoot, onecolumn, 12pt]{IEEEtran}

\usepackage{graphicx}

\begin{document}

\title{Non-Linear Temperature Dependence in  Graphene Nanoribbon Tunneling Transistors}

\author{Youngki~Yoon
        and~Sayeef~Salahuddin\\
        Department of Electrical Engineering and Computer Sciences\\
        University of California, Berkeley, CA 94720}


%
%
\maketitle
\begin{abstract}
It is usually assumed that tunneling current is fairly independent of temperature. By performing an atomistic transport simulation, we show, to the contrary, that the subthreshold tunneling current in a  graphene nanoribbon (GNR) band-to-band tunneling transistor (TFET) should show significant and non-linear temperature dependence. Furthermore, the nature of this non-linearity changes as a function of source/drain doping and vertical electric field, indicating that such non-linearity, if properly understood, may provide important insights into the tunneling phenomena. Finally, by developing a pseudo-analytical method, we predict that such temperature dependence is not unique to GNR but should rather be a general behavior for any band-to-band tunneling transistor independent of the channel material.
\end{abstract}
%
%
\newpage
\IEEEpeerreviewmaketitle
\section{Introduction}
Recently there has been significant interest in band-to-band
Tunneling Field Effect Transistors (TFET) due to the fact that
such devices may reduce power supply requirement by reducing the
subthreshold swing \cite{Bowonder08, Krishnamohan08,
Appenzeller04, Luisier09, Verhulst08, Koswatta09}. As a material,
graphene nanoribbon (GNR) is expected to be an attractive
candidate for TFETs due to low effective mass, narrow and direct
bandgap, and compatibility with planar processing
\cite{Zhang08,Zhao09}. In this letter, we explore the temperature
dependence of subthreshold current in a GNR TFET. Temperature
dependence in TFET structures has been rarely studied primarily
due to the traditional view that tunneling is independent of
temperature. This view stems from the fact that the Kane's model
\cite{Kane61}, that has been extensively used to interpret
experimental data \cite{Verhulst08,Motayed08}, presents an exact
solution only for zero temperature and thus does not include
effects of temperature exclusively. By performing an atomistic
simulation of GNR TFETs, we find, contrary to the traditional
view, that the subthreshold current in such a device should show
significant non-linearity with temperature. In one hand, this
non-linearity is very different in nature to the temperature
dependence of subthreshold current exhibited by a GNR FET. On the
other, it carries a distinct signature of the tunneling phenomena
itself that, if properly understood, may be utilized to give
important insights into the physics of band-to-band tunneling.
While our simulations were performed specifically for GNR TFETs,
by constructing a pseudo-analytical model we show that, such
non-linearity should be a generic feature of any TFET independent
of channel
 material.

\section{Approach}
For the channel and source/dain extension, we use an armchair-edge
Graphene Nanoribbon (aGNR) with $n = 13$ (width$\approx$1.6 nm)
that leads to an intrinsic bandgap of  $E_g$ = 0.86 eV with
H-terminated edges where edge bond relaxation is considered with a
modified tight-binding parameter ($t = 1.12 t_0$, $t_0 = -2.7$ eV)
\cite{Son06}. The nominal device has double-gate geometry with 1.6
nm thick $\rm{HfO_2}$ gate oxide ($\kappa$ = 16) resulting in an
equivalent oxide thickness EOT = 0.4 nm. The channel length is
$L_{ch}$ = 15 nm, and the source/drain extension is $L_{S/D}$ = 15
nm. Source/drain doping density, $N_{S/D}$ is set to $0.01
\rm{/atom}$, which is equivalent to $3.8\times10^{13} \rm{/cm}^2$.
A power supply voltage of $V_{DD} = 0.4$ V is assumed. The
electronic transport is modeled by solving the open boundary
Schr\"{o}dinger equation using non-equilibrium Green's function
(NEGF) formalism within the nearest neighbor tight-binding
approximation in a $\rm{p_z}$ orbital basis set (see for example
\cite{Zhao09} and references therein) that automatically accounts
for electron and hole currents. The NEGF transport equation is
solved self-consistently with a three-dimensional solution of
Poisson equation. For comparison, we explored two different types
of structures. Figure 1(a) shows energy band diagram of an n-i-n
MOSFET device that we shall use as a reference.  When the GNR is
p-doped at the source, the resulting p-i-n structure will
constitute a tunneling FET [Fig. 1(b)].

\section{Results}
$I_D-V_G$ characteristic for a GNR MOSFET is shown in Fig. 2(a)
for various temperatures. In general, the subthreshold swing of a MOSFET can
be written as $S = 60m \times {T}/{{300}}$, where $m$, often called the body factor, is a constant and equal
to the ratio of the supply voltage to the channel potential. Two
things can be easily inferred from this expression: First, the
thermionic emission current is expected to give a constant
subthreshold swing for a given temperature and second, the
subthreshold swing is expected to change linearly with
temperature. This is exactly what we see in Fig. 2(a).

By contrast, the $I_D - V_G$ characteristics of a TFET ([Fig.
2(b)]) show significant non-linearity in the subthreshold region
both over the voltage range and also over the temperature. In
absence of an analytical expression as in the case of MOSFETs, it
is difficult to understand the underlying physics in a simple
manner. In order to analyze the non-linear temperature dependence
in TFETs, we have plotted the shift of voltage, $\Delta V$ with
temperature at various current levels denoted as $I_{off}$ in Fig.
2(c). The $\Delta V$ is measured from $\Delta V = V^T  -
V^{300{\rm{ K}}}$ and it is fitted with the following equation:
\begin{eqnarray}
\label{eqn2}
\Delta V = A(300-T)^\gamma.
\end{eqnarray}
The rationale for using Eq. (\ref{eqn2}) is that the parameter
$\gamma$ gives information about non-linearity with temperature.
At the same time, any dependence of $\gamma$ with respect to
$I_{off}$ gives information about the non-linearity in the
subthreshold region as a function of gate voltage. Thus both
non-linearities in the tunneling current can be captured in a
single parameter $\gamma$. In Fig. 2(d), we show $\gamma$ as a
function of $I_{off}$. We see that, for MOSFET (the squares),
$\gamma$ is independent of $I_{off}$ and it is equal to 1. This
means that the subthreshold swing for a MOSFET is linear with
temperature and independent of gate voltage. On the other hand,
for the TFET (the circles), $\gamma$ is larger than 1 and it also
varies as a function of $I_{off}$. The non-linearity in $\gamma$
in Fig. 2(d) thus carries a telltale signature of the tunneling
phenomena itself, and experimental measurements can be
characterized in this way to distinguish between thermionic
emission type and band-to-band tunneling currents. Fig. 2(d) is
the central result of our paper.

To understand the non-linearity in $\gamma$, we note that, in the ballistic case, the current can be written as
\begin{eqnarray}
\label{eqn3}
I = \frac{{2e}}{h}\int {dE\bar T(E)\left\{ {f_1  -
f_2 } \right\}},
\end{eqnarray}
where $\bar T(E)$ is transmission, $h$ is the plank's constant and
$f_1$ and $f_2$ are the Fermi functions of the source and the
drain terminals, respectively. The inset in Fig. 2(b) shows the
transmission at $T = 300$ K as a function of energy, $E$ and gate
voltage, $V_G$. Taking this numerical result, we re-calculated
current from Eq. (\ref{eqn3}) at all other temperatures so that
the transmission is kept fixed for all temperatures and the
variation in temperature only enters in the Fermi functions. The
calculated currents are plotted as individual markers in Fig. 2(a)
and (b). We see that this pseudo-analytical treatment shows a
reasonably good agreement with the full numerical results for both
the MOSFET and the TFET structures. This clearly indicates that
any non-linearity that we see in the tunneling current as a
function of temperature is coming from $(f_1 - f_2)$. Using this fact, we can now construct a simple picture of the underlying physics. The band-diagram shown in Fig. 1(a) shows that in case of the MOSFET structure, the current has the full contribution of the Fermi tail. By contrast, as Fig. 1(b) shows, the Fermi tail is cut off by the bandgap for the TFET. This phenomenon of Fermi-tail cut-off is well known and is credited for the lowering of subthreshold swing in a tunnel transistor. What we show here is that this cut-off is also responsible for the non-linearity in the subthreshold swing with temperature and can be characterized as shown in Fig. 2(d).

Another very important conjecture from
Eq. (\ref{eqn3}) is its generality. No matter what material is
used, the temperature dependence is qualitatively going to be the
same since the specifics of the
difference in materials will only modify the transmission function
$\bar T(E,V_G )$. Strictly speaking, transmission should also be
affected by temperature through the charge self-consistency.
However, as long as the Fermi level is not right at the band edge
and the temperature is not extremely low, the temperature
dependence of $\bar T(E,V_G )$ is negligible.

Next we examine how this non-linearity is further influenced by
various device parameters. First we look at the equivalent gate
oxide thickness (EOT) that changes the vertical field. In our simulation this field is varied by changing the dielectric
constant of the insulator. What we see from Fig. 3(b) and (c) is
that the non-linearity in $\gamma$ increases significantly with
increasing EOT. To explain this behavior we plotted the band
profile of the TFET for EOT = 0.4 (solid line) and 0.8 nm (dashed
line) on the left panel of Fig. 3(a). It is seen that larger EOT
makes the tunneling barrier thinner since the gate loses its
control over the contacts right outside the channel region. A
thinner barrier effectively acts to cut less of the tail compared
to a thicker barrier as confirmed by the fact that the current
spectrum of a larger EOT exists at higher energy [dashed line on
the right panel in Fig. 3(a)]. It is important to note that a
larger EOT results in a thinner barrier and thus moves the effective
band-edge away from the Fermi level, thereby increasing the overall
non-linearlity. In turn, this increased non-linearity can carry
the information of EOT itself. Note that in comparison, the
$\Delta V$ and $\gamma$ for the MOSFET are unaffected by the
change of EOT [see the inset of Fig. 3(b) and dashed line in
3(c)].

Figure 3(d-f) show the effects of source/drain doping density on
the temperature dependence. Two observations can be made from the
band profiles [Fig. 3(d)]: Lower doping causes (i) a decrease in
the energy window for carrier injection, and (ii) an increase in
the thickness of the tunneling barrier. We see from Fig. 3(e) that
the voltage shift $\Delta V$ is drastically reduced with the low
doping density, which can again be understood from the Fermi tail
cut-off. A higher density moves the Fermi level away from the
band-edge, effectively lowering the amount of Fermi-tail
truncation by the band gap. As discussed earlier, this increases
the non-linearity in the current. On the other hand, a lower
density moves the Fermi level closer to the band-edge effectively
reducing the non-linearity. In order to quantify the effect of
doping on $\Delta V$ in TFETs, $\Delta V$ is plotted as a function
of $N_{S/D}$ in Fig. 3(f). For the simulated range, we see a
nearly linear dependence of $\Delta V$ with doping density. By
contrast, $\Delta V$ is almost independent of source/drain doping
for the MOSFET structure [see the inset in Fig. 3(f)]. Again here,
in addition to the fact that it is distinctly different from the
MOSFET current, the tunneling current in the TFET may provide
information about the source/drain doping density itself.

In this work, we have ignored phonon scattering that may bring in
its own temperature dependence in the tunneling current and will
have to be studied carefully. Nonetheless, this should merely
determine the exact degree of non-linearity in $\gamma$, keeping
the qualitative picture depicted in Fig. 2(d) intact. Note that,
even in presence of the aforementioned non-idealites, the MOSFET
retains a linear temperature dependence in subthreshold swing,
indicating a flat $\gamma$, and this fact has been verified by
many experiments, e.g., \cite{Michalas07}. Notably first
principles study suggests that change of bandgap with temperature
for graphene nanoribbons is minimal \cite{Son06}.

\section{Conclusion}
To summarize, based on an atomistic quantum transport simulation,
we predict that the band-to-band tunneling current in a GNR TFET
should exhibit unique non-linearity with temperature that carries
a distinct signature as a function of various device quantities
such as doping and vertical electric field, where MOSFET
characteristics remain unaffected. By constructing a
pseudo-analytical model to compare with our numerical results, we
have shown that these unique characteristics should be independent
of the channel material and generally applicable to any TFET. Thus
we believe that the temperature dependence can be used as a
generic spectroscopic tool that may give important insights into
band-to-band tunneling phenomena.
%
%
%
\newpage
\bibliographystyle{ieeetr}

\begin{thebibliography}{10}

\bibitem{Bowonder08}
A.~Bowonder et al., ``Low-voltage green transistor using ultra
shallow junction and hetero-tunneling'', {\em Int. Work. on Junc.
Tech. (IWJT)}, pp.~93-96, 2008.

\bibitem{Krishnamohan08}
T.~Krishnamohan et al.,  ``Double-Gate Strained-Ge Heterostructure
TFET with Record High Drive Current and less than 60 mV/dec
Subthreshold Slope'', {\em IEDM Tech. Dig.}, 2008.

\bibitem{Appenzeller04}
J.~Appenzeller et al., ``Band-to-Band Tunneling in Carbon Nanotube
Field-Effect Transistors'', {\em Phys. Rev. Lett.}, vol.~93,
no.~19, p.~196805, 2004.

\bibitem{Luisier09}
M.~Luisier et al., ``Atomistic Full-Band Design Study of InAs
Band-to-Band Tunneling Field-Effect Transistors'', {\em IEEE Elec.
Dev. Lett.}, vol.~30, no.~6, p.~602, 2009.

\bibitem{Verhulst08}
A.~S.~Verhulst et al., ``Boosting the on-current of a n-channel
nanowire tunnel field-effect transistor by source material
optimization'', {\em J. Appl. Phys.}, vol.~104, p.~064514, 2008.

\bibitem{Koswatta09}
S.~O. Koswatta et al., ``Performance Comparison Between p-i-n
Tunneling Transistors and Conventional MOSFETs'', {\em IEEE Trans.
Elec. Dev.}, vol.~56, no.~3, pp.~456-465, 2009.

\bibitem{Zhang08}
Q.~Zhang et al., ``Graphene Nanoribbon Tunnel Transistors'', {\em
IEEE Elec. Dev. Lett.}, vol.~29, no.~12, pp.~1344-1346, 2008.

\bibitem{Zhao09}
P.~Zhao et al., ``Computational Study of Tunneling Transistor
Based on Graphene Nanoribbon'', {\em Nano Lett.}, vol.~9, no.~2,
pp.~684-688, 2009.

\bibitem{Kane61}
E.O.~Kane, ``Theory of Tunneling'', {\em J. Appl. Phys.}, vol.~32,
no.~1, p.~83, 1961.

\bibitem{Motayed08}
A.~Motayeda et al., ``GaN-nanowire/amorphous-Si core-shell
heterojunction diodes'', {\em Appl. Phys. Lett.}, vol.~93,
p.~193102, 2008.

\bibitem{Son06}
Y.~W. Son et al., ``Energy gaps in graphene nanoribbons'', {\em
Phys. Rev. Lett.}, vol.~97, no.~21, p.~216803, 2006.

\bibitem{Michalas07}
L.~Michalas et al., ``An experimental study of the thermally
activated processes in polycrystalline silicon thin film
transistors'', {\em Microelectronics Reliability}, vol.~47,
no.~12, pp.~2058-2064, 2007.
\end{thebibliography}

\newpage
\begin{figure}[b]
\centering
\includegraphics[scale=0.8]{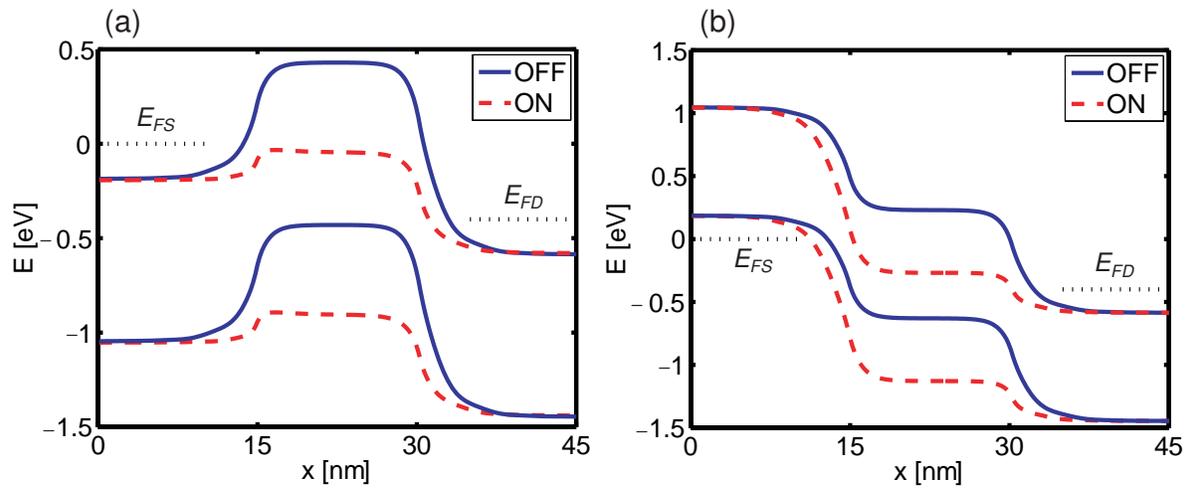}
\caption{Band diagram of GNR (a) MOSFET and (b) TFET at the on
(dashed line) and the off (solid line) states.} \label{fig1}
\end{figure}

\newpage
\begin{figure}[t]
\centering
\includegraphics[scale=0.8]{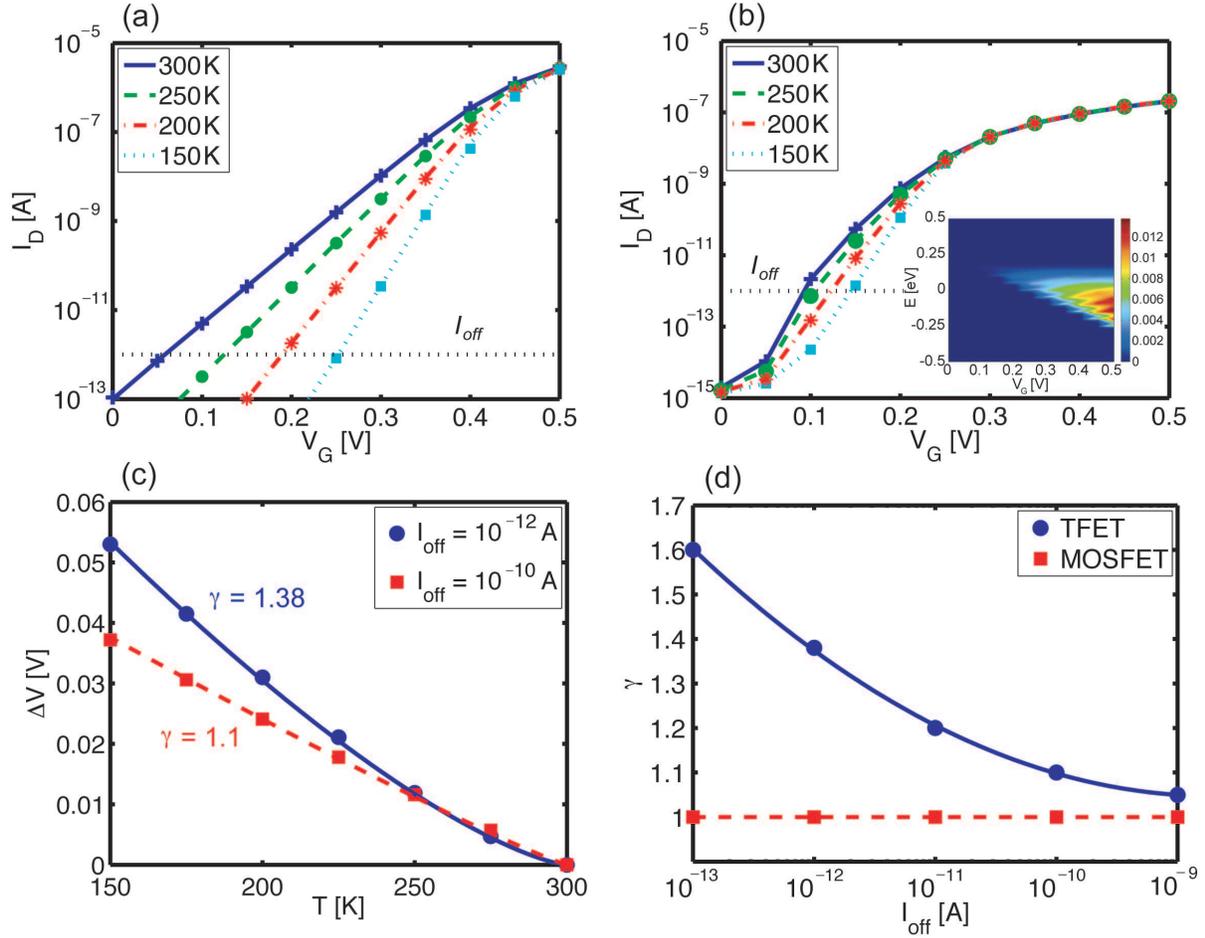}
\caption{$I_D - V_G$ characteristics of GNR (a) MOSFET and (b)
TFET; the solid, the dashed, the dash-dot, the dotted lines are
for $T$ = $300$, $250$, $200$, $150$ K, respectively. The markers
are calculated from Eq. (\ref{eqn3}) with $\bar T(E, V_G)$ at 300
K (inset). (c) The voltage shift, $\Delta V$ at a common current
level, read from Fig. 2(b). Lines are plotted using Eq.
(\ref{eqn2}) for fitting. (d) Variation of $\gamma$ as a function
of $I_{off}$, clearly showing the difference in behavior of the
tunneling current in comparison to the thermionic current. Note
that GNR TFETs show an ambipolar behavior. Here we assumed gate
metal work function difference, $\Phi _{ms} = qV_D/2$ so that the
minimal leakage current is achieved at $V_G$ = 0 \cite{Zhao09},
which is the symmetric point of ambipolar conduction.}
 \label{fig2}
\end{figure}

\newpage
\begin{figure}[t]
\centering
\includegraphics[scale=0.7]{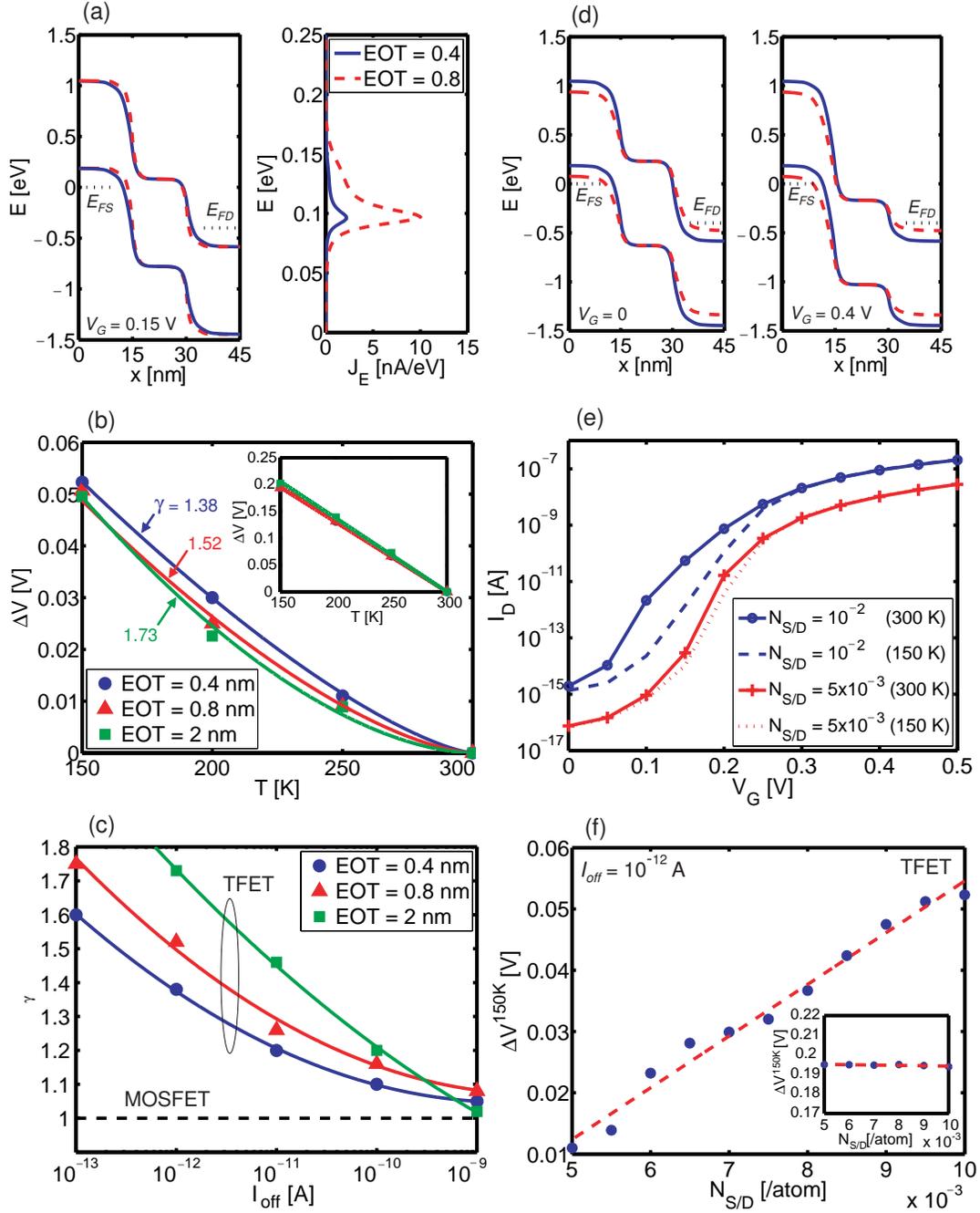}
\caption{(a) Band diagram (left panel) and energy-resolved current
spectrum (right panel) for equivalent oxide thickness, EOT = 0.4
(solid line) and 0.8 nm (dashed line) in TFETs at $V_G$ = 0.15 V.
(b) The voltage shift $\Delta V$ vs. $T$ for various EOTs at
$I_{off} = 10^{-12}$ A in TFETs (main panel) and in MOSFETs
(inset). (c) Variation of $\gamma$ as a function of $I_{off}$ for
different EOTs. (d) Band diagram for $N_{S/D} = 10^{-2}$ (solid
line) and $5 \times 10^{-3} \rm{/atom}$ (dashed line) in TFETs at
$V_G$ = 0 (left) and 0.4 V (right). (e) $I_D - V_G$
characteristics for two different doping and temperature showing
the extent of voltage shift with temperature as a function of
doping. (f) The voltage shift $\Delta V$ as a function of doping
density. $\Delta V$ shows a roughly linear dependence with doping
for TFETs (main panel), but is almost independent of doping for
MOSFET (inset).} \label{fig3}
\end{figure}

\end{document}